\documentstyle[12pt]{article}
\def\etal{et al.\ }

\def\bold#1{\setbox0=\hbox{$#1$}%
     \kern-.025em\copy0\kern-\wd0
     \kern.05em\copy0\kern-\wd0
     \kern-.025em\raise.0433em\box0 }

\begin{document}

\baselineskip 18pt

\newcommand{\sheptitle}
{Radiative Corrections to Charged Higgs Production
\\ in
${\bold{e^+e^-}}$ Colliders${}^{\*}$}

\newcommand{\shauthor}
{Marco A. D\'\i az}

\newcommand{\shaddress}
{Physics Department, University of Southampton\\
Southampton, SO17 1BJ, U.K.}

\newcommand{\vauthor}
{Tonnis A. ter Veldhuis}
\newcommand{\vaddress}
{Department of Physics and Astronomy, Vanderbilt University\\
Nashville, TN 37235, USA}

\newcommand{\shepabstract}
{We study one loop electroweak corrections
to the production of a pair of charged Higgs bosons through an
intermediate $Z$-boson or photon. In particular, we
consider the effects of graphs with top and bottom quarks and squarks
in the loop within the context of the Minimal Supersymmetric Model.
We find that the corrections can be considerable, and typically are
of the order of 10\% to 20\%.}

\begin{titlepage}
\begin{flushright}
VAND-TH-94-25\\
SHEP 95-02\\
January 1995 \\
\end{flushright}
\vspace{.4in}
\begin{center}
{\large{\bf \sheptitle}\footnotemark}
\bigskip \\ \shauthor \\ \mbox{} {\it \shaddress} \\ \vspace*{0.3cm}
and
\bigskip \\ \vauthor \\ \mbox{} {\it \vaddress} \\
{\bf Abstract} \bigskip \end{center} \setcounter{page}{0}
\shepabstract

\footnotetext{Presented by Tonnis ter Veldhuis at the
Eighth Meeting of the Division of Particles and Fields of the American
Physical Society \lq\lq DPF94'', The University of New Mexico, Albuquerque,
August 2-6, 1994, and at ``Beyond the Standard Model IV'',
Granlibakken, Lake Tahoe, December 13-18, 1994.}

\end{titlepage}

\newpage

The minimal supersymmetric standard model (MSSM) \cite{MSSMrep}
provides a very interesting extension of the standard model.
The absence of quadratic divergencies in this model is an immediate
consequence of its supersymmetric nature.
The ensuing improved UV behavior ameliorates the technical
aspect of the naturalness problem
that haunts the standard model (SM). Furthermore, it is tempting
to believe that the unification of
the gauge coupling constants is not coincidental \cite{alphauni};
the required SUSY breaking scale is small
enough to avoid the reintroduction of the naturalness problem, and the
unification scale is large enough to suppress proton decay yet
smaller than the Planck scale. Indeed, a
supergravity theory with non-perturbative SUSY breaking may even
explain the hierarchy of energy scales.

The Higgs sector of the MSSM contains two neutral Higgs bosons,
a pseudo scalar Higgs boson,
and a pair of charged Higgs bosons \cite{hhg}.
The next generation of $e^+e^-$ colliders, notably LEP2
and the NLC, will be able to probe the MSSM Higgs sector
and test the model.
It is therefore important to understand the production and decay
of charged Higgs bosons in detail.

Radiative corrections in the MSSM can be considerable because
of the large mass splitting between the top and
bottom quark. For instance,
the tree-level upper bound on the lightest Higgs mass is $M_Z$, but
radiative corrections increase this upper bound to $130$ GeV \cite{nhm}.
Here we study the one-loop electroweak corrections to the
production cross-section of a pair of charged Higgs bosons in
$e^+e^-$ colliders.

We work in an on-shell scheme \cite{diaz}, in which $M_Z^2 = 1/4 (g^2+{g'}^2)
v^2$
and $M_W^2= 1/4 g^2 v^2$, with $v^2=v_t^2+v_b^2$, are defined to be
the physical masses of the vector bosons.
Similarly, we define the tree-level expression
for the mass of the pseudoscalar in terms of the
renormalized parameters to be the physical mass.
The residues of the poles  of the photon and the pseudoscalar propagators
are equal to one in our scheme, and the mixing between the photon and the
Z boson vanishes at zero momentum.
The normalization
conditions that complete the set  relevant to this paper are
imposed on the coupling of the pseudoscalar to charged leptons
\begin{equation}
\Gamma_{Al^+l^-}|_{p^2=M_A^2} = - \frac{g m_l \tan \beta}{2 M_W},
\end{equation}
and the coupling of the photon to electrons
\begin{equation}
\Gamma_{\gamma e^+e^-}^{\mu}|_{p^2=0} = i e \gamma^{\mu}.
\end{equation}
The angles $\theta_W$ and $\beta$ are defined by
$\sin^2 \theta_W  = 1 - M_W^2/M_Z^2$ and
$\tan \beta  =  v_b/v_t$,
and the electric charge is $e=g g' /\sqrt{g^2 +{g'}^2}$.

We consider the production of a pair of on-shell charged Higgs bosons.
The momenta $k_\mu$ and $k_{\mu}'$ of the outgoing charged Higgs bosons
therefore satisfy $k^2={k'}^2=M_{H^{\pm}}^2$.
The amplitude for the process is given by
\begin{eqnarray}
I & = & -\frac{g^2 \cos 2 \theta_W}{4 \cos^2 \theta_W}{\bar v}_{p_2}
\gamma_{\mu} (v_z-a_z \gamma_5) u_{p_1}
\frac{g^{\mu \nu}}{p^2-M_Z^2} (f_Z k_{\nu} -f_{Z}'k_{\nu}')- \nonumber \\
 & & e^2 {\bar v}_{p_2} \gamma_{\mu} (v_{\gamma}-a_{\gamma}\gamma_5) u_{p_1}
\frac{g^{\mu \nu}}{p^2} (f_{\gamma} k_{\nu} -f_{\gamma}'k_{\nu}').
\end{eqnarray}
Here $v_z=1/2-2\sin^2 \theta_W$,
$a_z=1/2$,$v_{\gamma}=1$ and $a_{\gamma}=0$.
The form factors $f_Z$, $f_Z'$, $f_{\gamma}$ and $f_{\gamma}'$ are
equal to one at tree-level. Radiative
corrections manifest themselves in modifications of these form factors
\begin{eqnarray}
f_Z & = & 1 + \frac{{\tilde A}_{ZZ}}{p^2-M_Z^2} +
\tan 2\theta_W \frac{{\tilde A}_{\gamma Z}}{p^2} + {\tilde A}_{ZH^+H^-}
+{\tilde A}_{H^+H^-}^d (M_{H^{\pm}}), \nonumber \\
f_Z' & = & 1 + \frac{{\tilde A}_{ZZ}}{p^2-M_Z^2} +
\tan 2\theta_W \frac{{\tilde A}_{\gamma Z}}{p^2} + {\tilde A}_{ZH^+H^-}'
+{\tilde A}_{H^+H^-}^d (M_{H^{\pm}}), \nonumber \\
f_{\gamma} & = & 1 + \frac{{\tilde A}_{\gamma \gamma }}{p^2} +
\frac{1}{{\tan 2\theta}_W} \frac{{\tilde A}_{Z \gamma}}{p^2-M_Z^2} +
{\tilde A}_{\gamma H^+H^-}+{\tilde A}_{H^+H^-}^d (M_{H^{\pm}}), \nonumber \\
f_{\gamma}' & = & 1 + \frac{{\tilde A}_{\gamma \gamma}}{p^2} +
\frac{1}{{\tan 2\theta}_W} \frac{{\tilde A}_{Z \gamma}}{p^2-M_Z^2} +
{\tilde A}_{\gamma H^+H^-}'+{\tilde A}_{H^+H^-}^d (M_{H^{\pm}}) ,
\end{eqnarray}
where the tilde indicates renormalized quantities with infinities subtracted
in accordance with
the normalization conditions.
The various terms in the modified form factors are related to
the gauge boson self-energies,
${\tilde \Pi}_{ij}^{\mu\nu}=i{\tilde A_{ij}} g^{\mu\nu} + i{\tilde B_{ij}}
p^{\mu} p^{\nu}$
with $i,j=\gamma,Z$,
to the loop contributions of the vertices,
${\tilde \Lambda}_{ZH^+H^-}^{\mu}=i{\tilde A}_{ZH^+H^-}k^{\mu}
- i{\tilde A}_{ZH^+H^-}'{k'}^{\mu}$
and
${\tilde \Lambda}_{\gamma H^+H^-}^{\mu}={\tilde A}_{\gamma H^+H^-}k^{\mu}
- {\tilde A}_{\gamma H^+H^-}'{k'}^{\mu}$,
and to the charged Higgs boson self-energy,
$i{\tilde A}_{H^+H^-}^d = \frac{d}{dp^2} {\tilde \Sigma}_{H^+H^-}$.
We calculate one-loop oblique and vertex corrections.
As large radiative corrections are expected to arise from
the high value of the top quark mass, we furthermore limit our
calculation to loops with top and bottom quarks, and with
stop and sbottom squarks.
In this approximation there are no contributions from box diagrams.
The differential cross-section therefore has the
characteristic $\sin^2 \theta$
scattering angle dependence of s-channel processes.
The total cross-section in terms of the form factors is
\begin{eqnarray}
\sigma
& = & \frac{ \left(1 - 4 M_{H^{\pm}}/s \right)^{\frac{3}{2}}}
{48 \pi s }
\left\{ \frac{g^4 \cos^2 2 \theta_w}{16 \cos^4 \theta_w}
\frac{{v_z}^2 + {a_z}^2}{\left(1-M_Z^2/s\right)^2}
\left| \frac{f_Z+f_Z'}{2}\right|^2 + \right. \nonumber \\
 &  & e^4 \left| \frac{f_{\gamma}+f_{\gamma}'}{2} \right|^2 +
\left.
2 \frac{g^2 \cos 2 \theta_w}{4 \cos^2 \theta_w} e^2 \frac{v_z}
{\left(1-M_Z^2/s \right)} \Re \left[
\left( \frac{f_Z +f_Z'}{2} \right)\left(\frac{f_{\gamma} + f_{\gamma}'}{2}
\right)^{*}\right]
\right\}
\end{eqnarray}
where $\sqrt{s}$ is the center of mass energy, and $M_{H^{\pm}}$ is the
renormalized charged Higgs boson mass \cite{chm}.

Radiative corrections enter the total cross-section both through
corrections to the charged Higgs boson mass $M_{H^{\pm}}$ and through
corrections to the form factors.
Corrections to the charged higgs boson mass $M_{H^{\pm}}$ manifest
themselves in the fact that the tree-level mass relation
$M_{H^{\pm}}^2 = M_W^2 + M_A^2$ is not longer valid.
As shown in Fig.\ 1, departures from this mass relation are
most pronounced for very small and very large values of $\tan \beta$.

In Fig.\ 2a  we plot the total cross-section as a function of the
center of mass energy $\sqrt{s}$
for various values of $\tan \beta$
and the pseudoscalar mass $M_A$.
We plot the tree-level cross-section (dashed lines) and the
cross-section including all one-loop corrections (solid lines)
within our approximations.
All soft SUSY breaking squark mass
parameters are equal to $M_{SUSY}=400\ GeV$, and the Peccei-Quinn symmetry
breaking parameter $\mu$ as well as the soft SUSY breaking trilinear
couplings are taken to be zero, a choice that implies no mixing
in the squark sector. The top quark mass is taken to be
$m_t =174\ GeV$, the central value of the recent CDF result \cite{CDF}.
Because the curves in Fig.\ 2a reflect a fixed
value of the pseudoscalar higgs boson mass $M_A$, radiative corrections
to the charged Higgs boson mass cause a shift in the production threshold;
to lower
$\sqrt{s}$ for low values of $\tan \beta$, and to higher $\sqrt{s}$
for high values of $\tan \beta$. For intermediate values
of $\tan \beta$ the dashed and the dot-dashed curves almost
coincide, even near threshold, reflecting
the negligible radiative corrections to the charged Higgs boson mass in this
range. The solid curve corresponding to
$\tan\beta=0.5$ has a discontinuity in slope at
$\sqrt{s}=2m_t=348$ GeV. This discontinuity stems from
the triangular graph with two top quarks
and a bottom quark in the internal lines,
and its effect is enhanced at low values of $\tan\beta$.
In Fig.\ 2b we again show the cross-section as a function of the
center of mass energy $\sqrt{s}$, but now for fixed values
of the charged higgs boson mass $M_{H^{\pm}}$. Accordingly
the production threshold at tree-level (dashed line) and the
at one loop (solid line) coincide, but the values of the pseudoscalar higgs
mass $M_A$ at tree and one-loop level differ.
For center of mass energies well above threshold, the solid
lines show that for our choice of parameters
one-loop corrections to the form factors
reduce the cross-section by 10\% to 25\% with respect to
the tree-level cross-section.
{}.

In Fig.\ 3 we plot the ratio between the one-loop renormalized
cross section and the tree level cross section, as a function
of $\tan\beta$ for various values of the top quark mass for fixed
values of the pseudoscalar mass $M_A$
(dashed lines). To separate the effect of the corrections to
the charged Higgs mass, we plot the ratio between the
one-loop renormalized cross section and the cross section with
corrected $M_{H^{\pm}}$, but form factors equal to one
(solid lines). Solid and dashed lines coincide at intermediate
values of $\tan\beta$ because in that region the corrections to
the charged Higgs mass are negligible,
as can be seen in Fig.\ 1,
where the renormalized charged Higgs mass is shown as
a function of $\tan\beta$ for the same values of the top quark
mass.
We appreciate from Fig.\ 3 that the corrections to the cross
section are typically between -10\% to -20\% in the intermediate
region of $\tan\beta$ for our choice of parameters.
For more extreme values of $\tan\beta$
corrections can be larger and can have either sign. These large
corrections stem mainly from corrections to the form factors.
\\

\noindent
{\bf Note:} While we were completing this work we received a preprint
concerning a similar calculation \cite{Arhrib}.

\vskip 1cm

\vskip 1cm

\noindent
{\bf \large Figure Captions:}

\noindent
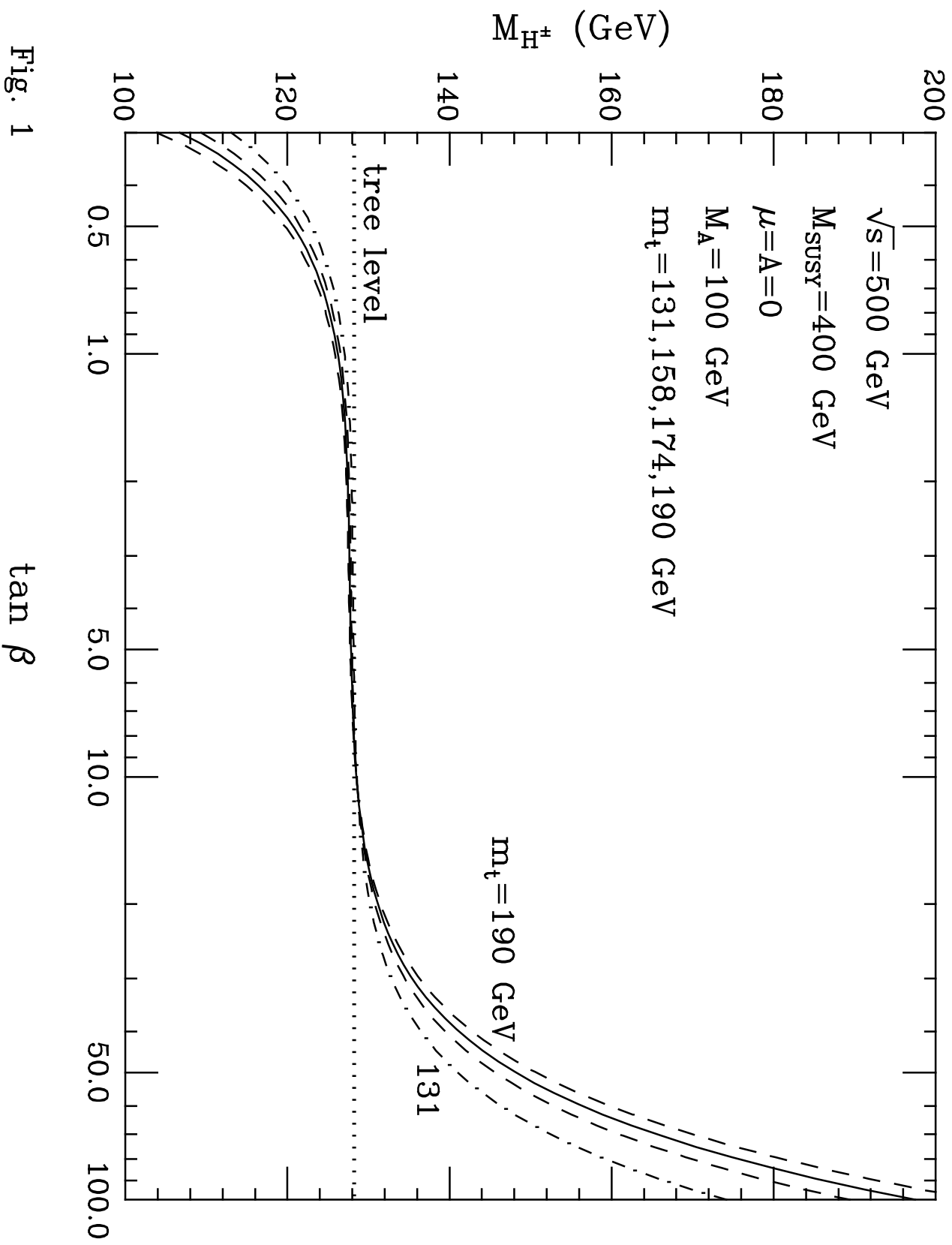
{\bf Fig. 1.}
Renormalized charged Higgs mass as a function of $\tan\beta$
for different values of the top quark mass. In solid we plot the
CDF central value $m_t=174$ GeV. In dashes we have the $1\sigma$
deviations from the central value: $m_t=158$ and 190 GeV. In
dotdashes we plot the $95\%$ c.l. lower limit $m_t=131$ GeV.
The horizontal dotted line corresponds to the tree level charged
Higgs mass.\\
{\bf Fig. 2a.}
The cross-section $\sigma(e^+ e^- \rightarrow H^+ H^-)$,
as a function of the center of mass energy $\sqrt{s}$
for fixed values of $M_A$. Dashed (solid) lines
show the tree-level (one-loop) result. \\
{\bf Fig. 2b.}
The cross-section $\sigma(e^+ e^- \rightarrow H^+ H^-)$,
as a function of the center of mass energy $\sqrt{s}$
for fixed values of $M_{H^{\pm}}$. Dashed (solid) lines
show the tree-level (one-loop) result. \\
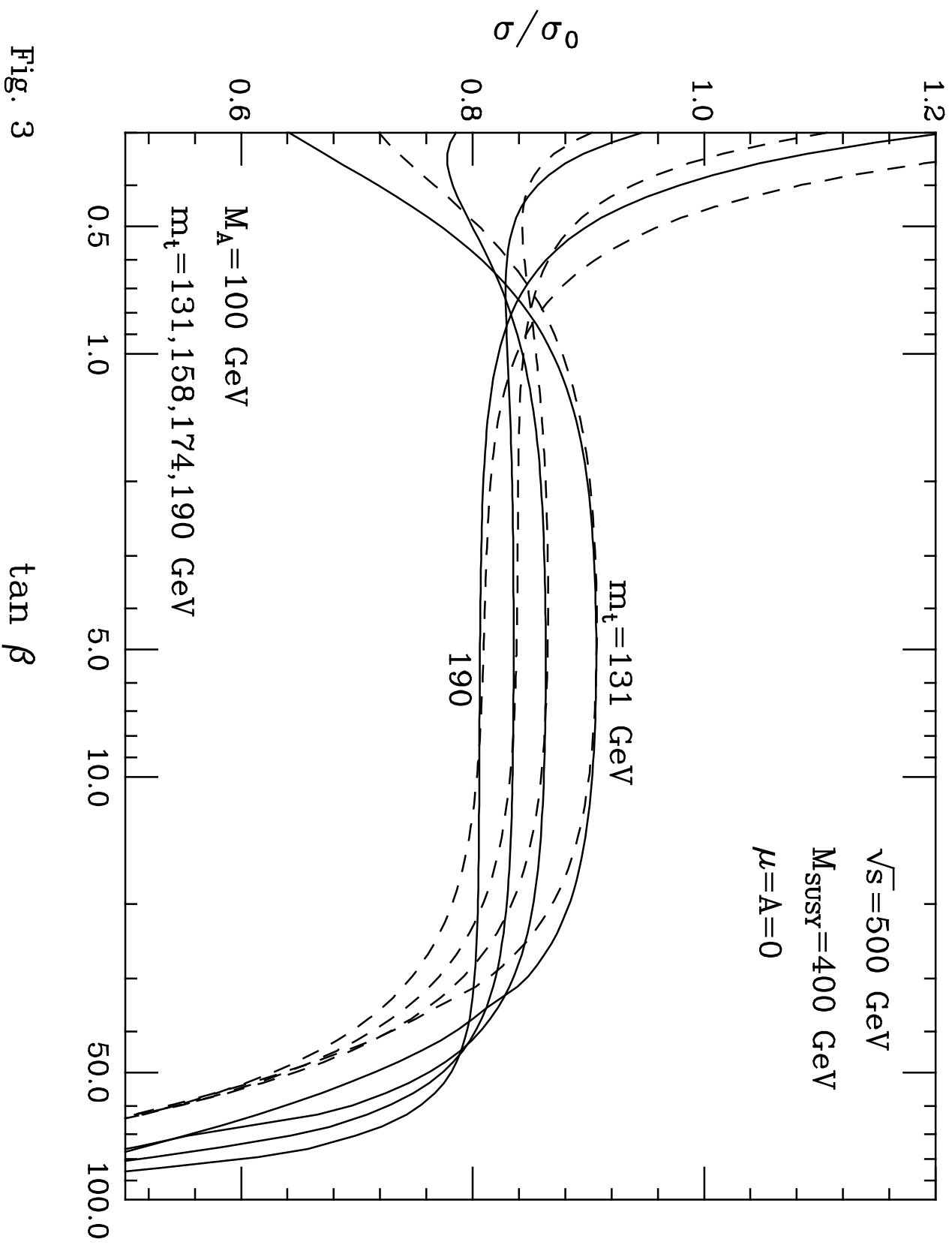
{\bf Fig. 3.}
Ratio between the one-loop renormalized cross section and the
tree level cross section as a function of $\tan\beta$ for different
values of the top quark mass (dashed lines). For comparison
we also plot the ratio between the one-loop renormalized cross
section and the cross section with corrected charged Higgs mass
but form factors equal to one (solid lines).\\

\end{document}